\documentclass{my-elsarticle}

\usepackage{a4wide}
\usepackage{amssymb,amsmath,amsthm}
\usepackage{epsfig,graphicx,epstopdf}
\usepackage{hyperref}

\allowdisplaybreaks

\newtheorem{theorem}{Theorem}

\theoremstyle{definition}
\newtheorem{definition}{Definition}[theorem]

\begin{document}
	
\begin{frontmatter}
		
\title{Mathematical Modeling of COVID-19 Transmission Dynamics\\ with a Case Study of Wuhan}

\author[Add:a,Add:b]{Fa\"{\i}\c{c}al Nda\"{\i}rou}
\ead{faical@ua.pt}

\author[Add:b]{Iv\'an Area}
\ead{area@uvigo.es}

\author[Add:c]{Juan J. Nieto}
\ead{juanjose.nieto.roig@usc.es}

\author[Add:a]{Delfim F. M. Torres\corref{corD}}
\ead{delfim@ua.pt}
\cortext[corD]{Corresponding author.}

\address[Add:a]{Center for Research and Development in Mathematics and Applications (CIDMA),\\ 
Department of Mathematics, University of Aveiro, 3810-193 Aveiro, Portugal}

\address[Add:b]{Departamento de Matem\'atica Aplicada II, 
E. E. Aeron\'autica e do Espazo, Campus de Ourense,\\ 
Universidade de Vigo, 32004 Ourense, Spain} 

\address[Add:c]{Instituto de Matematicas, Universidade de Santiago de Compostela, 
15782 Santiago de Compostela, Spain}


\begin{abstract}
We propose a compartmental mathematical model for the spread of 
the COVID-19 disease with special focus on the transmissibility 
of super-spreaders individuals. We compute the basic reproduction 
number threshold, we study the local stability of the disease free equilibrium
in terms of the basic reproduction number, and we investigate 
the sensitivity of the model with respect to the variation 
of each one of its parameters. Numerical simulations 
show the suitability of the proposed COVID-19 model 
for the outbreak that occurred in Wuhan, China.
\end{abstract}


\end{frontmatter}


\section{Introduction}

Mathematical models of infectious disease transmission dynamics are now ubiquitous. 
Such models play an important role in helping to quantify possible infectious disease 
control and mitigation strategies \cite{MR3808514,MR3897752,MR3544685}. There exist 
a number of models for infectious diseases; as for compartmental models, starting 
from the very classical SIR model to more complex proposals \cite{Brauer}.

Coronavirus disease 2019 (COVID-19) is an infectious disease caused by severe acute 
respiratory syndrome coronavirus 2 (SARS-CoV-2). The disease was first identified 
December 2019 in Wuhan, the capital of Hubei, China, and has since spread globally, 
resulting in the ongoing 2020 pandemic outbreak \cite{worldometer}. 
The COVID-19 pandemic is considered as the biggest global threat worldwide 
because of thousands of confirmed infections, accompanied by thousands deaths 
over the world. Notice, by March 26, 2020, report 503,274 confirmed cumulative 
cases with 22,342 deaths. At the time of this revision, the numbers have increased 
to 1,353,361 confirmed cumulative cases with 79,235 deaths, according to the report 
dated by April 8, 2020, by the Word Health Organization.

The global problem of the outbreak has attracted the interest of researchers 
of different areas, giving rise to a number of proposals to analyze and predict 
the evolution of the pandemic \cite{Chen,Maier}. Our main contribution is related 
with considering the class of super-spreaders, which is now appearing in medical 
journals (see, e.g., \cite{Trilla,wong}). This new class, as added to any 
compartmental model, implies a number of analysis about disease 
free equilibrium points, which is also considered in this work.

The manuscript is organized as follows. In Section~\ref{sec:model}, 
we propose a new model for COVID-19. A qualitative analysis 
of the model is investigated in Section~\ref{sec:qam}:
in Section~\ref{sec:R0}, we compute the 
basic reproduction number $R_0$ of the COVID-19 system model;
in Section~\ref{subsec:Stab}, we study the local stability 
of the disease free equilibrium in terms of $R_0$.
The sensitivity of the basic reproduction number $R_0$ 
with respect to the parameters of the system model
is given in Section~\ref{sec:sensitivity}.
The usefulness of our model is then illustrated in Section~\ref{sec:numerical}  
of numerical simulations, where we use real data from Wuhan. 
We end with Section~\ref{sec:conc} of conclusions, 
discussion, and future research.


\section{The Proposed COVID-19 Compartment Model}
\label{sec:model}

\begin{figure}[!htb]
\centering 
\includegraphics[scale=0.75]{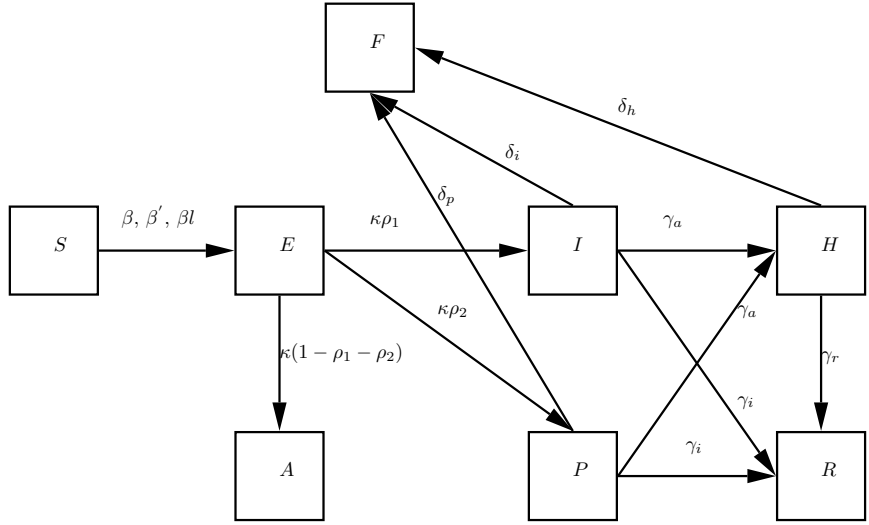}
\caption{Flowchart of model \eqref{model}.}
\label{figure:flowchart}
\end{figure}

Based on a 2016 model \cite{kim}, and taking into account the existence 
of super-spreaders in the family of corona virus \cite{hussain}, 
we propose a new epidemiological compartment model that takes 
into account the super-spreading phenomenon of some individuals. 
Moreover, we consider a fatality compartment, related to death due 
to the virus infection. In doing so, the constant total population 
size $N$ is subdivided into eight epidemiological classes: 
susceptible class ($S$), exposed class ($E$), symptomatic and infectious class ($I$), 
super-spreaders class ($P$), infectious but asymptomatic class ($A$), hospitalized ($H$), 
recovery class ($R$), and fatality class ($F$). The model takes the following form:
\begin{equation}
\label{model}
\begin{cases}
\displaystyle{\frac{d S}{dt} = -\beta\frac{I}{N}S-l\beta \frac{H}{N}S-\beta^{'}\frac{P}{N}S},\\[3mm]
\displaystyle{\frac{d E}{dt}= \beta\frac{I}{N}S+l\beta \frac{H}{N}S+ \beta^{'}\frac{P}{N}S -\kappa E}, \\[3mm]
\displaystyle{\frac{d I}{dt}= \kappa \rho_1 E - (\gamma_a + \gamma_i)I-\delta_i I}, \\[3mm]
\displaystyle{\frac{d P}{dt}= \kappa \rho_2 E- (\gamma_a + \gamma_i)P-\delta_p P}, \\[3mm]
\displaystyle{\frac{d A}{dt}= \kappa (1-\rho_1 - \rho_2)E },\\[3mm]
\displaystyle{\frac{d H}{dt}= \gamma_a (I + P) - \gamma_r H - \delta_h H}, \\[3mm]
\displaystyle{\frac{d R}{dt}= \gamma_i (I + P)+ \gamma_r H},\\[3mm]
\displaystyle{\frac{dF}{dt}= \delta_i I(t) + \delta_p P(t) + \delta_h H(t)},
\end{cases}
\end{equation}
with $\beta$ quantifying the human-to-human transmission 
coefficient per unit time (days) per person, 
$\beta^{'}$ quantifies a high transmission coefficient due to super-spreaders, 
and $l$ quantifies the relative transmissibility of hospitalized patients. 
Here $\kappa$ is the rate at which an individual leaves the exposed class 
by becoming infectious (symptomatic, super-spreaders or asymptomatic); 
$\rho_1$ is the proportion of progression from exposed class $E$ to symptomatic 
infectious class $I$; $\rho_2$ is a relative very low rate at which exposed individuals 
become super-spreaders while $1-\rho_1-\rho_2$ is the progression 
from exposed to asymptomatic class; $\gamma_a$ is the average
rate at which symptomatic  and super-spreaders individuals become hospitalized; 
$\gamma_i$ is the recovery rate without being hospitalized; $\gamma_r$ 
is the recovery rate of hospitalized patients; and $\delta_i$, $\delta_p$, 
and $\delta_h$ are the disease induced death rates due to infected, 
super-spreaders, and hospitalized individuals, respectively.
At each instant of time,
\begin{equation}
\label{model2}
D(t) := \delta_i I(t) + \delta_p P(t) + \delta_h H(t) = \frac{dF(t)}{dt}
\end{equation}
gives the number of death due to the disease. The transmissibility from asymptomatic 
individuals has been modeled in this way since it was not apparent their behavior. 
Indeed, at present, this question is a controversial issue for epidemiologists. 
A flowchart of model \eqref{model} is presented in Figure~\ref{figure:flowchart}.

\section{Qualitative Analysis of the Model}
\label{sec:qam}

One of the most significant thresholds when studying infectious disease models,
which quantifies disease invasion or extinction in a population, 
is the basic reproduction number \cite{van2017reproduction}. 
In this section we obtain the basic reproduction number for our model \eqref{model}
and study the locally asymptotically stability of its disease free equilibrium
(see Theorem~\ref{thm01}).


\subsection{The Basic Reproduction Number}
\label{sec:R0}

The basic reproduction number, as a measure for disease spread in a population, 
plays an important role in the course and control of an ongoing outbreak. 
It can be understood as the average number of cases one infected individual 
generates, over the course of its infectious period, 
in an otherwise uninfected population. Using the next generation matrix 
approach outlined in \cite{van} to our model \eqref{model}, 
the basic reproduction number can be computed by
considering the below generation matrices $F$ and $V$, that is, 
the Jacobian matrices associated to the rate of appearance of new 
infections and the net rate out of the corresponding compartments, respectively,
\begin{equation*}
J_{\mathcal{F}}
= \left[ \begin{array}{cccc}
0& \beta & \beta l & \beta^{'}\\
0& 0 & 0&0\\
0&0& 0&0\\
0&0& 0&0
\end{array}
\right] \quad \text{ and } \quad
J_{\mathcal{V}}
= \left[ \begin{array}{cccc}
\kappa & 0 & 0&0\\
-\kappa \rho_1& \varpi_i &0&0\\
-\kappa \rho_2& 0& \varpi_p &0\\
0& -\gamma_a&-\gamma_a&\varpi_h
\end{array}
\right],
\end{equation*}
where 
\begin{gather}
\label{varpi}
\varpi_i = \gamma_a + \gamma_i + \delta_i,\,\,
\varpi_p = \gamma_a + \gamma_i + \delta_p 
\text{ and } \varpi_h =\gamma_r + \delta_h .
\end{gather}
The basic reproduction number $R_0$ is obtained 
as the spectral radius of $F \cdot V^{-1}$, precisely,
\begin{equation}
\label{R0}
R_0= \frac{\beta \rho_1(\gamma_a l + \varpi_h)}{\varpi_i  \varpi_h} 
+ \frac{(\beta \gamma_a l + \beta^{'}\varpi_h) \rho_2}{\varpi_p \varpi_h}.
\end{equation}
For the parameters used in our simulations (see Table~\ref{tab:param}), 
one computes this basic reproduction number to obtain $R_0=0.945$. 
This means that the epidemic outbreak that has occurred in Wuhan 
was well controlled by the Chinese authorities.


\subsection{Local Stability in Terms of the Basic Reproduction Number}
\label{subsec:Stab}

Noting that the two last equations and the fifth of system \eqref{model} 
are uncoupled to the remaining equations of the system, 
we can easily obtain, by direct integration, the following analytical results:
\begin{equation}
\begin{cases}
A(t)= \kappa (1-\rho_1 -\rho_2)\int^t_0 E(s)ds\\[3mm]
R(t)= \gamma_i \int^t_0\Big( I(s)+ P(s)\Big)ds + \gamma_r \int^t_0 H(s)ds\\[3mm]
F(t)= \delta_i\int^t_0 I(s)ds + \delta_p\int^t_0 P(s)ds + \delta_h\int^t_0 H(s)ds.
\end{cases}
\end{equation}
Furthermore, since the total population size $N$ is constant, one has
\begin{equation}
S(t)= N-\left[ E(t) + I(t) + P(t) + A(t) + H(t) + R(t)+ F(t) \right].
\end{equation}
Therefore, the local stability of model \eqref{model} 
can be studied through the remaining coupled system of states variables, 
namely, the variables $E$, $I$, $P$, and $H$ in \eqref{model}. 
The Jacobian matrix associated to these variables 
of \eqref{model} is the following one:
\begin{equation}
J_M= \left[ \begin{array}{cccc}
-\kappa & \beta & l\beta & \beta^{'}\\
\kappa \rho_1& -\varpi_i &0 &0\\
\kappa \rho_2& 0& -\varpi_p &0\\
0& \gamma_a&\gamma_a&-\varpi_h
\end{array}
\right],
\end{equation}
where $\varpi_i$, $\varpi_p$, and $\varpi_h$ 
are defined in \eqref{varpi}.
The eigenvalues of the matrix $J_M$ are the roots 
of the following characteristic polynomial:
\[
Z(\lambda)= \lambda^4 + a_1\lambda^3 + a_2\lambda^2 + a_1\lambda + a_4,
\]
where
\begin{align*}
a_1&= \kappa + \varpi_h + \varpi_i + \varpi_p,\\[3mm]
a_2&= -\beta \kappa \rho_1 -\beta^{'}\kappa \rho_2 
+ \kappa\varpi_h + \kappa \varpi_i + \varpi_h \varpi_i 
+ \kappa \varpi_p + \varpi_h \varpi_p + \varpi_i \varpi_p,\\[3mm]
a_3&= -\beta \gamma_a \kappa l\rho_1 -\beta \gamma_a \kappa l\rho_2 
- \beta \kappa\rho_1 \varpi_h - \beta^{'} \kappa\rho_2 \varpi_h 
- \beta \kappa\rho_1 \varpi_p - \beta^{'} \kappa\rho_2 \varpi_i \\
&\qquad + \kappa\varpi_h\varpi_i + \kappa\varpi_h\varpi_p 
+ \kappa\varpi_i\varpi_p+ \varpi_h\varpi_i\varpi_p,\\[3mm]
a_4&= -\beta \gamma_a \kappa l \rho_2\varpi_i -\beta \gamma_a 
\kappa l \rho_1\varpi_p - \beta^{'} \kappa \rho_2\varpi_i\varpi_h 
- \beta \kappa \rho_1\varpi_h \varpi_p + \kappa \varpi_h\varpi_i\varpi_p.
\end{align*}
Next, by using the Li\'enard--Chipard test \cite{Gant,Lie}, 
all the roots of $Z(\lambda)$ are negative or have negative 
real part if, and only if, the following conditions are satisfied:
\begin{itemize}
\item[1.]  $a_i>0$, $i= 1, 2, 3, 4$;
\item[2.] $a_1 a_2 > a_3$.
\end{itemize}
In order to check these conditions of the Li\'enard--Chipard test, 
we rewrite the coefficients $a_1$, $a_2$, $a_3$, and $a_4$ 
of the characteristic polynomial in terms of the basic 
reproduction number given by \eqref{R0}:
\begin{eqnarray*}
a_1&=& \kappa + \varpi_h + \varpi_i + \varpi_p,\\[3mm]
a_2&=& (1-R_0)(\kappa \varpi_i + \kappa \varpi_p) 
+ \kappa \varpi_p\frac{\beta\rho_1}{\varpi_i} 
+ \kappa \varpi_i\frac{\beta^{'}\rho_2}{\varpi_p} 
+ \beta\gamma_al\rho_1\kappa\left( \frac{1}{\varpi_h} 
+ \frac{\varpi_p}{\varpi_h \varpi_i}\right) \\[3mm]
&& + \beta\gamma_al\rho_2\kappa\left( \frac{1}{\varpi_h} 
+ \frac{\varpi_i}{\varpi_h \varpi_p}\right) + (\kappa + \varpi_i)
\varpi_h + (\varpi_h + \varpi_i)\varpi_p,\\[3mm]
a_3&=& \kappa(1-R_0)(\varpi_h\varpi_p + \varpi_h\varpi_i 
+ \varpi_i\varpi_p) + \kappa\varpi_p\frac{\beta\rho_1\varpi_h}{\varpi_i} 
+ \kappa\varpi_i\frac{\beta^{'}\rho_2\varpi_h}{\varpi_p} \\[3mm]
&&+ \kappa\varpi_p\beta\gamma_al\rho_1\left( \frac{1}{\varpi_h} 
+ \frac{1}{\varpi_i}\right) + \kappa\varpi_i\beta\gamma_al\rho_2\left( 
\frac{1}{\varpi_h} + \frac{1}{\varpi_p}\right) + \varpi_i\varpi_h\varpi_p,\\[3mm]
a_4&=&\kappa\varpi_i\varpi_h\varpi_p(1-R_0) .
\end{eqnarray*}
Moreover, we also compute, in terms of $R_0$, the following expression:
\begin{align*}
a_1a_2 -a_3&= (1-R_0)(\kappa + \varpi_i)\kappa\varpi_i 
+ (1-R_0)(\kappa + \varpi_h + \varpi_p)\kappa\varpi_p\\[3mm]
& + (\kappa + \varpi_p + \varpi_i)\Big(\frac{\beta\rho_1}{\varpi_p} 
+ \frac{\beta \gamma_al\rho_1}{\varpi_i}\Big)\kappa\varpi_p
+ (\kappa + \varpi_p + \varpi_i)\Big(\frac{\beta^{'}\rho_2}{\varpi_p}
+  \frac{\beta\gamma_al\rho_2}{\varpi_p}\Big)\kappa\varpi_i \\[3mm]
& + (\kappa + \varpi_h + \varpi_i)\frac{\beta\gamma_al\rho_1\kappa}{\varpi_h}  
+ (\kappa + \varpi_h + \varpi_p)\frac{\beta\gamma_al\rho_2\kappa}{\varpi_h}  
+ (\kappa + \varpi_i)\varpi_h + (\varpi_h + \varpi_i)\varpi_p.
\end{align*}
From these previous expressions, it is clear that if $R_0 <1$, then
the conditions of the Li\'enard--Chipard test are satisfied and, 
as a consequence, the disease free equilibrium is stable.
In the case when $R_0 > 1$, we have that $a_4 < 0$ and, 
by using Descartes' rule of signs, we conclude that at least 
one of the eigenvalues is positive. Therefore, the system is unstable.
In conclusion, we have just proved the following result:

\begin{theorem}
\label{thm01}
The disease free equilibrium of system \eqref{model}, that is, $(N,0,0,0,0,0,0)$,  
is locally asymptotically stable if $R_0 < 1$ and unstable if $R_0 > 1$. 
\end{theorem}

Next we investigate the sensitiveness of the COVID-19 model \eqref{model},
with respect to the variation of each one of its parameters, 
for the endemic threshold \eqref{R0}.


\section{Sensitivity Analysis}
\label{sec:sensitivity}

As we saw in Section~\ref{sec:qam}, the basic reproduction number 
for the COVID-19 model \eqref{model}, which we propose
in Section~\ref{sec:model}, is given by \eqref{R0}.
The sensitivity analysis for the endemic threshold \eqref{R0}
tells us how important each parameter is to disease transmission.
This information is crucial not only for experimental design,
but also to data assimilation and reduction of complex models
\cite{powell2005sensitivity}. Sensitivity analysis is commonly used
to determine the robustness of model predictions to parameter values,
since there are usually errors in collected data and presumed parameter values.
It is used to discover parameters that have a high impact on the threshold $R_0$
and should be targeted by intervention strategies. More accurately,
sensitivity indices' allows us to measure the relative change
in a variable when a parameter changes. For that purpose, we use the normalized
forward sensitivity index of a variable with respect to a given parameter,
which is defined as the ratio of the relative change in the variable
to the relative change in the parameter. If such variable is differentiable
with respect to the parameter, then the sensitivity index is defined as follows.

\begin{definition}[See \cite{chitnis2008determining,rodrigues2013sensitivity}]
\label{def:sentInd}
The normalized forward sensitivity index of $R_0$, which is differentiable
with respect to a given parameter $\theta$, is defined by
$$
\Upsilon_\theta^{R_0}=\frac{\partial R_0}{\partial \theta}\frac{\theta}{R_0}.
$$
\end{definition}

The values of the sensitivity indices for the parameters values of Table~\ref{tab:param},
are presented in Table~\ref{tab:sensitivity}. 

\begin{table}[ht!]
\centering
\caption{Values of the model parameters corresponding to the situation 
of Wuhan, as discussed in Section~\ref{sec:numerical}, for which $R_0 = 0.945$.  }
\begin{tabular}{|l|l|l| l |} \hline
Name & Description & Value & Units \\ \hline
$\beta$ &  Transmission coefficient from infected individuals  & 2.55 & day$^{-1}$\\
$l$ & Relative  transmissibility of hospitalized patients& $1.56$ & dimensionless\\
$\beta^{'}$ & Transmission coefficient due to super-spreaders & 7.65 & day$^{-1}$\\
$\kappa$ & Rate at which exposed become infectious & 0.25 & day$^{-1}$\\
$\rho_1$ & Rate at which exposed people become infected $I$ & 0.580 & dimensionless\\
$\rho_2$ & Rate at which exposed people become super-spreaders  & 0.001 & dimensionless\\
$\gamma_a$  & Rate of being hospitalized & 0.94 & day$^{-1}$\\
$\gamma_i$ & Recovery rate without being hospitalized & 0.27 & day$^{-1}$\\
$\gamma_r$ & Recovery rate of hospitalized patients & 0.5 & day$^{-1}$\\
$\delta_i$& Disease induced death rate due to infected class & 3.5 & day$^{-1}$\\
$\delta_p$& Disease induced death rate due to super-spreaders & 1 & day$^{-1}$\\
$\delta_h $ & Disease induced death rate due to  hospitalized class & 0.3 & day$^{-1}$\\ \hline
\end{tabular}
\label{tab:param}
\end{table}
\begin{table}[!htb]
\centering
\caption{Sensitivity of $R_0$ evaluated for the parameter 
values given in Table~\ref{tab:param}.}\label{tab:sensitivity}
\begin{tabular}{|c|c|} \hline
Parameter & Sensitivity index \\[1mm] \hline
$\beta$ &   0.963\\[1mm]
$l$ & 0.631\\[1mm]
$\beta^{'}$ & 0.366\\[1mm]
$\kappa$ & 0000\\[1mm]
$\rho_1$ &  0.941\\[1mm]
$\rho_2$ &  0.059\\[1mm]
$\gamma_a$  &  0.418\\[1mm]
$\gamma_i$ &  -0.061\\[1mm]
$\gamma_r$ &  -0.395\\[1mm]
$\delta_i$&  -0.699\\[1mm]
$\delta_p$& -0.027\\[1mm]
$\delta_h $ & -0.238\\[1mm] \hline
\end{tabular}
\end{table}

These values have been determined experimentally in such a way 
the mathematical model describes well the real data, giving rise to 
Figures~\ref{fig:Inf} and \ref{fig:death}. 
Other values for the parameters can be found, e.g., in \cite{aguilar}.

Note that the sensitivity index may depend on several parameters of the system, 
but also can be constant, independent of any parameter. For example, 
$\Upsilon_{\theta}^{R_0}=+1$ means that increasing (decreasing) 
$\theta$ by a given percentage increases (decreases) always $R_0$ by that same percentage. 
The estimation of a sensitive parameter should be carefully done, since a small perturbation 
in such parameter leads to relevant quantitative changes. On the other hand, 
the estimation of a parameter with a rather small value for the sensitivity index 
does not require as much attention to estimate, because a small 
perturbation in that parameter leads to small changes.

From Table~\ref{tab:sensitivity}, we conclude that the most sensitive parameters 
to the basic reproduction number $R_0$ of the COVID-19 model \eqref{model}
are $\beta$, $\rho_1$ and $\delta_i$. 
In concrete, an increase of the value of $\beta$ will increase 
the basic reproduction number by $96.3\%$ and this happens, 
in a similar way, for the parameter $\rho_1$. In contrast, 
an increase of the value of $\delta_i$ will decrease $R_0$ by $69.9\%$.


\section{Numerical Simulations: The Case Study of Wuhan}
\label{sec:numerical}

We perform numerical simulations to compare the results of our model 
with the real data obtained from several reports published by WHO 
\cite{who1,who2} and worldometer \cite{worldometer}.

The starting point of our simulations is 4 January 2020 (day 0), 
when the Chinese authorities informed about the new virus \cite{who1}, 
with already 6 confirmed cases in one day. From this period up to January 19, 
there is less information about the number of people contracting the disease. 
Only on January 20, we have the report \cite{who2}, with 1460 new reported 
cases in that day and 26 the dead. Thus, the infection gained much more attention 
from 21 January 2020, with 1739 confirmed cases and 38 the dead,
up to 4 March 2020, when the numbers in that day were as 
low as 11 and 7, respectively infected and dead,
after a pick of 3892 confirmed cases on 27 January 2020 
and a pick of 254 dead on 4 February 2020.
Here we follow the data of the daily reports published by \cite{worldometer}.
We show that our COVID-19 model describes well the real data of daily confirmed cases
during the 2 months outbreak (66 days to be precise, from January 4 to March 9, 2020).

The total population of Wuhan is about 11 million. During the COVID-19 outbreak, 
there was a restriction of movements of individuals due to quarantine in the city. 
As a consequence, there was a limitation on the spread of the disease. 
In agreement, in our model we consider, as the total population under study, 
$N= 11000000/250$. This denominator has been determined in the first days 
of the outbreak and later has been proved to be a correct value: 
according to the real data published by the WHO, it is an appropriate value 
for the restriction of movements of individuals. As for the initial conditions, 
the following values have been fixed: $S_0=N-6$, $E_0=0$, $I_0=1$, 
$P_0=5$, $A_0=0$, $H_0=0$, $R_0=0$, and $F_0=0$.

\begin{figure}[!htb]
\centering \label{Inf:data}
\includegraphics[scale=0.5]{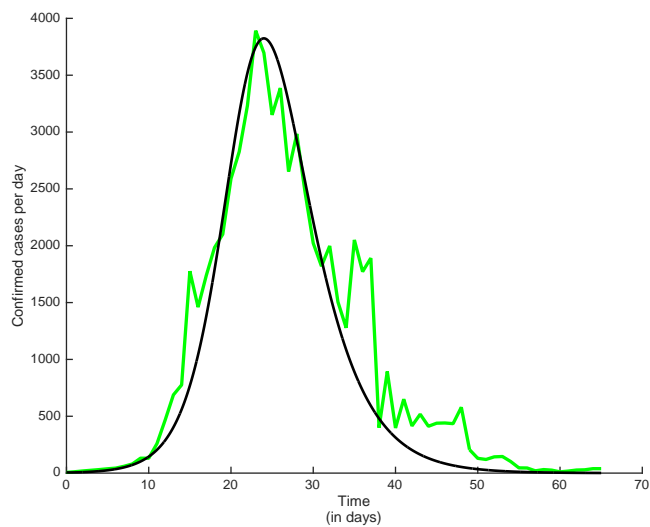}
\caption{Number of confirmed cases per day. The green line corresponds 
to the real data obtained from reports \cite{who1,who2,worldometer} 
while the black line ($I + P + H  $) has been obtained by solving numerically 
the system of ordinary differential equations \eqref{model}, 
by using the Matlab code \texttt{ode45}.}
\label{fig:Inf}
\end{figure}

We would like to mention that there exist gaps in the reports of the WHO 
at the beginning of the outbreak. For completeness, we give here the list
$L_C$ of the number of confirmed cases in Wuhan per day, corresponding
to the green line of Figure~\ref{fig:Inf}, and the list
$L_D$ of the number of dead individuals in Wuhan per day, corresponding
to the red line of Figure~\ref{fig:death}:
\begin{multline*}
L_C = 
[6,
12,
19,
25,
31,
38,
44,
60,
80,
131,
131,
259,
467,
688,
776,
1776,
1460,
1739,
1984,
2101,
2590,
2827,\\
3233,
3892,
3697,
3151,
3387,
2653,
2984,
2473,
2022,
1820,
1998,
1506,
1278,
2051,
1772,
1891,
399,
894,\\
397,
650,
415,
518,
412,
439,
441,
435,
579,
206,
130,
120,
143,
146,
102,
46,
45,
20,
31,
26,
11,
18,
27,
29,
39,
39],
\end{multline*}
\begin{multline*}
L_D = [
0,
0,
0,
0,
0,
0,
0,
0,
4,
4,
4,
8,
15,
15,
25,
26,
26,
38,
43,
46,
45,
57,
64,
66,
73,
73,
86,
89,
97,
108,
97,
254,\\
121,
121,
142,
106,
106,
98,
115,
118,
109,
97,
150,
71,
52,
29,
44,
37,
35,
42,
31,
38,
31,
30,
28,
27,
23,
17,
22,
11,\\
7,
14,
10,
14,
13,
13
].
\end{multline*}
Lists $L_C$ and $L_D$ have 66 numbers, where $L_C(0)$ represents the number of confirmed cases 
04 January 2020 (day 0) and $L_C(65)$ the number of confirmed cases 09 March 2020 (day 65)
and, analogously, $L_D(0)$ represents the number of dead on January 4 and $L_D(65)$ the number
of dead on March 9, 2020.


\section{Conclusions and Discussion}
\label{sec:conc}

\begin{figure}[!htb]
\centering
\label{Cum:death:data}\includegraphics[scale=0.5]{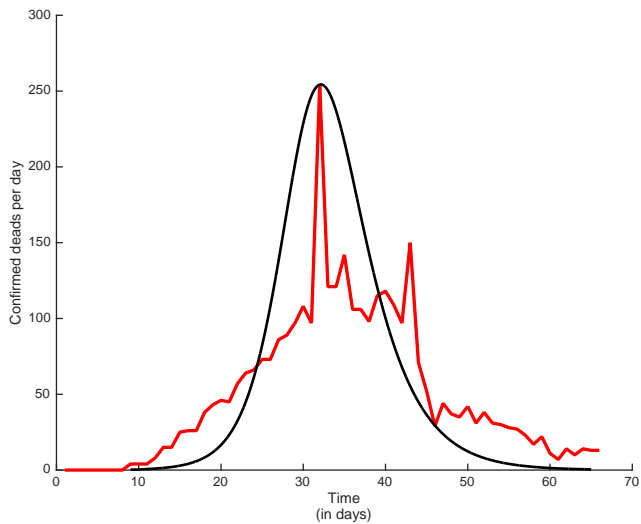}
\caption{Number of confirmed deaths per day. The red line corresponds 
to the real data obtained from reports \cite{who1,who2,worldometer} 
while the black line has been obtained by solving numerically,
using the Matlab code \texttt{ode45}, our system of ordinary 
differential equations \eqref{model} to derive $D(t)$ 
given in \eqref{model2}.}
\label{fig:death}
\end{figure}

Classical models consider SIR populations. Here we have taken into consideration 
the super-spreaders ($P$), hospitalized ($H$), and fatality class ($F$), 
so that its derivative (see formula \eqref{model2}) gives the number of deaths ($D$). 
Our model is an {\emph{ad hoc}} compartmental model of the COVID-19,
taking into account its particularities, some of them still not well-known, 
giving a good approximation of the reality of the Wuhan outbreak 
(see Figure~\ref{fig:Inf}) and predicting a diminishing on the daily number 
of confirmed cases of the disease. This is in agreement with our computations 
of the basic reproduction number in Section~\ref{R0} that, surprisingly, 
is obtained less than 1. Moreover, it is worth to mention that
our model fits also enough well the real data of daily 
confirmed deaths, as shown in Figure~\ref{fig:death}. 

Our theoretical findings and numerical results adapt well to the real 
data and it reflects or reflected the reality in Wuhan, China.
The number of hospitalized persons is relevant to give an estimate 
of the Intensive Care Units (ICU) needed. Some preliminary simulations 
indicate that this would be useful for the health authorities.
Our model can also be used to study the reality 
of other countries, whose outbreaks are currently on the rise. 
We claim that some mathematical models like the one we have proposed here 
will contribute to reveal some important aspects of this pandemia.

Of course, this investigation has some limitations, being the first 
on the relative recent spread of the new coronavirus and therefore the 
limited data accessible at the beginning of this study. In the future, 
we can develop further this prototype. Even with these shortcomings, 
the model can be useful due to the high relevance of the topic.
Finally, we suggest new directions for further research:
\begin{enumerate}
\item the transmissibility from asymptomatic individuals;

\item compare, in the near future, our results with other models;

\item consider sub-populations related to age, gender, etc.;

\item introduce preventive measures in this COVID-19 epidemic and for future viruses;

\item integrate into the model some imprecise data by using fuzzy differential equations;

\item include the viral load of the infectious into the model.
\end{enumerate}
These and other questions are under current investigation 
and will be addressed elsewhere.


\section*{Funding}

This research was funded by the
Portuguese Foundation for Science and Technology (FCT)
within project UIDB/04106/2020 (CIDMA).
Nda\"{\i}rou is also grateful to the support 
of FCT through the Ph.D. fellowship PD/BD/150273/2019. 
The work of Area and Nieto has been partially supported by the 
Agencia Estatal de Investigaci\'on (AEI) of Spain, cofinanced 
by the European Fund for Regional Development (FEDER) corresponding 
to the 2014-2020 multiyear financial framework, project MTM2016-75140-P.  
Moreover, Nieto also thanks partial financial support 
by Xunta de Galicia under grant ED431C 2019/02.



\end{document}